\documentclass[conference, a4paper]{IEEEtran}
\IEEEoverridecommandlockouts
\usepackage{cite}
\usepackage{amsmath,amssymb,amsfonts}
\usepackage{algorithmic}
\usepackage{graphicx}
\usepackage{textcomp}
\usepackage{xcolor}
\usepackage{booktabs}
\usepackage{stmaryrd}
\usepackage{comment}
\usepackage{tabularx}
\usepackage{url}
\usepackage{colortbl}
\definecolor{lightgray}{rgb}{0.6, 0.8, 1}
\usepackage{enumerate}
\usepackage{mathtools}
\usepackage{multirow}
\usepackage[hidelinks]{hyperref}
\usepackage{svg}
\usepackage{tikz}
\usetikzlibrary{arrows.meta, positioning, shapes, calc}
\usepackage{float}
\usepackage[table]{xcolor}
\usepackage{makecell}
\usepackage{adjustbox}
\usepackage{threeparttable}

\usepackage{tikz}
\usepackage{eso-pic}

\usetikzlibrary{arrows.meta,positioning,calc}
\definecolor{seedblue}{RGB}{28,67,103}
\definecolor{pathorange}{RGB}{230,105,20}
\definecolor{leafgreen}{RGB}{237,247,243}

\def\BibTeX{{\rm B\kern-.05em{\sc i\kern-.025em b}\kern-.08em
    T\kern-.1667em\lower.7ex\hbox{E}\kern-.125emX}}
\begin{document}

\title{Efficient TCitH-Based Alternatives to SLH-DSA: Cross-Layer ASIC Design of Mirath\\
\thanks{This work was partly funded by the German Federal Ministry of Research,
Technology and Space as part of the project ``PoQ-KiKi'' under grant
number 16KIS2064.}}

\AddToShipoutPictureFG*{%
\begin{tikzpicture}[remember picture,overlay]
\node[
anchor=south,
align=center,
text width=0.90\paperwidth,
font=\footnotesize
] at ([yshift=5mm]current page.south) {%
\copyright~2026 IEEE. Personal use of this material is permitted.
Permission from IEEE must be obtained for all other uses, in any current
or future media, including reprinting/republishing this material for
advertising or promotional purposes, creating new collective works,
for resale or redistribution to servers or lists, or reuse of any
copyrighted component of this work in other works.
};
\end{tikzpicture}%
}

\author{
\IEEEauthorblockN{Hiandra Tomasi, Maximilian Schöffel, Johannes Feldmann, and Norbert Wehn}
\IEEEauthorblockA{\textit{Microelectronic Systems Design Research Group} \\
\textit{RPTU Kaiserslautern-Landau}\\
Kaiserslautern, Germany\\
\{tomasi, m.schoeffel, j.feldmann, norbert.wehn\}@rptu.de} 
}

\maketitle

\begin{abstract}

To address the security risks posed by quantum computers, the U.S. National Institute of Standards and Technology (NIST) has standardized the post-quantum signature schemes ML-DSA, FN-DSA, and SLH-DSA. While ML-DSA and FN-DSA are lattice-based, SLH-DSA relies on hash-based assumptions. To support cryptographic agility against future vulnerabilities, NIST is evaluating non-lattice candidates as alternatives to SLH-DSA. Among these, TCitH-based schemes are particularly promising due to their compact keys and small signatures.

However, their high computational complexity and memory footprint pose significant challenges for efficient implementations on resource-constrained embedded platforms. 
They remain largely unexplored in this context, particularly in ASIC implementations.
To address this gap, we use a cross-layer methodology combining algorithmic and hardware layers to present, to the best of our knowledge, the first ASIC implementation of Mirath, a TCitH-based signature scheme, in a RISC-V-based system.
The design is implemented in a 22\,nm FD-SOI technology node.
Compared with an SLH-DSA ASIC implemented in the same technology node, the proposed architecture achieves 17.8$\times$ lower signing latency while requiring 58\,\% less total cell area, showing the potential of TCitH-based signatures as efficient non-lattice alternatives from an implementation perspective.

\end{abstract}


\section{Introduction}

The threat of quantum computers (QCs)~\cite{mosca2025global} has accelerated the transition to Post-Quantum Cryptography (PQC). The National Institute of Standards and Technology (NIST) selected ML-DSA, FN-DSA, and SLH-DSA for standardization~\cite{NISTFIPS2042024, FouqueEtAlFalcon2022, NISTFIPS2052024} and subsequently launched an additional digital-signature standardization process~\cite{NISTAdditionalSignatures2023} to strengthen crypto-agility, i.e., the ability to replace algorithms if vulnerabilities emerge. To justify adoption, non-lattice candidates are required to provide a significant performance advantage over SPHINCS+, the scheme underlying SLH-DSA~\cite{NISTAdditionalSignatures2023}.

Multi-Party Computation-in-the-Head (MPCitH) signatures~\cite{IshaiEtAlZeroKnowledgeMPC2007} are promising candidates in this process. In particular, the Threshold-Computation-in-the-Head (TCitH) framework~\cite{FeneuilRivainTCitH2025}, adopted by Mirath~\cite{AdjEtAlMirath2025}, RYDE~\cite{AragonEtAlRYDE2025}, and MQOM~\cite{BenadjilaEtAlMQOM2025}, significantly reduces signature sizes relative to earlier corresponding MPCitH-based designs. These schemes combine small public keys and compact signatures with security assumptions distinct from those of the standardized schemes. However, their reference implementations are computationally and memory intensive, leading to high latency, memory usage, and energy consumption on resource-constrained devices.

Efficient implementations of MPCitH-based schemes remain comparatively under-explored, and some candidates were eliminated before optimized implementation baselines were established~\cite{AlagicEtAlNISTIR8610_2026}. The state-of-the-art (SoA) consists mainly of software optimizations for off-the-shelf embedded devices~\cite{benadjila2026breaking, aranha2025faest, bettaieb2024enabling} and FPGA implementations~\cite{deshpande2024sdith, funk2026hake, schoffel2025hw}, while ASIC implementations of schemes from this process remain largely unexplored. In contrast, ASIC implementations of standardized PQC schemes are well documented~\cite{karl2025performance, saarinen2024sloth, dolmeta2026horcrux, karl2024riscv, carril2026pqcuark}.

In this context, hardware/software (HW/SW) co-design based on the RISC-V Instruction Set Architecture (ISA)~\cite{RISCVISA2026} is an established approach for embedded PQC~\cite{saarinen2024sloth,karl2024riscv,karl2025performance,dolmeta2026horcrux,carril2026pqcuark}, combining software programmability with dedicated acceleration of computational bottlenecks. Following this approach, we present a RISC-V-based HW/SW co-design of Mirath to challenge the perception that MPCitH-based signatures are inherently inefficient~\cite{KannwischerEtAlPQM42024}. We use Mirath as a representative TCitH-based scheme to investigate their computational and memory bottlenecks.

The main contributions of this work are:
\begin{enumerate}
    \item We present, to the best of our knowledge, the first ASIC implementation of a TCitH-based signature scheme. Our design is implemented in 22\,nm FD-SOI technology.

    \item We profile Mirath to identify computational and memory bottlenecks, develop dedicated accelerators for dominant computational kernels, and perform a HW/SW design-space exploration to quantify the individual and combined impact of the proposed accelerators on execution time.

    \item From a crypto-agility perspective, we demonstrate TCitH-based signatures as an efficient non-lattice alternative, achieving lower signing latency than all considered SoA SPHINCS+/SLH-DSA ASIC implementations and execution times comparable to SoA Dilithium/ML-DSA and Falcon/FN-DSA implementations.
\end{enumerate}

This paper is organized as follows: Section II provides the theoretical background. Section III describes the design objectives and cross-layer methodology. Section IV details the algorithmic optimizations, while Section V presents the hardware implementation. Experimental results and comparisons to SoA are provided in Sections VI and VII, respectively, followed by the conclusion in Section VIII.

\section{Background}
\begin{figure}[!t]
\centering
\resizebox{\columnwidth}{!}{%
\begin{tikzpicture}[
    font=\scriptsize,
    >=Latex,
    box/.style={
        draw,
        rounded corners,
        align=left,
        thick,
        inner sep=4pt,
        text width=2.35cm,
        minimum height=1.0cm
    },
    title/.style={
        draw,
        fill=gray!15,
        rounded corners,
        align=center,
        thick,
        inner sep=3pt,
        minimum width=2.35cm,
        minimum height=0.65cm
    },
    arrow/.style={->, thick},
    darrow/.style={->, dashed, thick}
]

% horizontal positions
\def\xA{0}
\def\xB{3.05}
\def\xC{6.10}

% vertical positions
\def\ytop{0}
\def\yone{-1.60}
\def\ytwo{-3.85}
\def\ythree{-5.85}

% -------------------------
% Titles
% -------------------------
\node[title] (kgtitle) at (\xA,\ytop) {\textbf{KeyGen}};
\node[title] (sgtitle) at (\xB,\ytop) {\textbf{Sign}};
\node[title] (vftitle) at (\xC,\ytop) {\textbf{Verify}};

% -------------------------
% KeyGen
% -------------------------
\node[box] (kg1) at (\xA,\yone) {
\textbf{K1: Sample and Expand seeds}\\
Sample and expand $\mathrm{seed}_{sk},\mathrm{seed}_{pk}$ to obtain
$\boldsymbol{S},\boldsymbol{C}'$ and $\boldsymbol{H}'$.
};

\node[box] (kg2) at (\xA,\ytwo) {
\textbf{K2: Compute syndrome $y$}\\
Build $\boldsymbol{E}=(\boldsymbol{S}\mid\boldsymbol{S}\boldsymbol{C}')$,
set $\mathbf{e}=\mathrm{vec}(\boldsymbol{E})$, and compute $\boldsymbol{y}$.
};

\node[box] (kg3) at (\xA,\ythree) {
\textbf{K3: Output keys}\\
$pk=(\mathrm{seed}_{pk},\boldsymbol{y})$\\
$sk=(\mathrm{seed}_{sk},\mathrm{seed}_{pk})$
};

\draw[arrow] (kgtitle) -- (kg1);
\draw[arrow] (kg1) -- (kg2);
\draw[arrow] (kg2) -- (kg3);

% -------------------------
% Sign
% -------------------------
\node[box] (sg1) at (\xB,\yone) {
\textbf{S1: Commit witnesses}\\
Derive randomized witness shares,
and compute commitments with digest $h_{\mathrm{com}}$.
};

\node[box] (sg2) at (\xB,\ytwo) {
\textbf{S2: Compute proof}\\
Derive $h_{\mathrm{sh}}$ and $\boldsymbol{\Gamma}$, compute
the polynomial-proof values $\boldsymbol{\alpha}^{(e)}_{\mathrm{mid}}$, $\boldsymbol{\alpha}^{(e)}_{\mathrm{base}}$, and $h_{\mathrm{piop}}$.
};

\node[box] (sg3) at (\xB,\ythree) {
\textbf{S3: Open and output}\\

Compute $\pi_{\mathrm{BAVC}}$ and output the signature $\sigma$.
};
%$\sigma=(\mathrm{salt},\mathrm{ctr},h_{\mathrm{piop}},\pi_{\mathrm{BAVC}},\ldots)$.
\draw[arrow] (sgtitle) -- (sg1);
\draw[arrow] (sg1) -- (sg2);
\draw[arrow] (sg2) -- (sg3);

% -------------------------
% Verify
% -------------------------
\node[box] (vf1) at (\xC,\yone) {
\textbf{V1: Reconstruct}\\
Parse $\sigma$, recover the opened shares,
rebuild commitments, and recompute $h_{\mathrm{sh}}$.
};

\node[box] (vf2) at (\xC,\ytwo) {
\textbf{V2: Recompute proof}\\
Derive $\boldsymbol{\Gamma}$ and recompute the evaluation/base
proof terms and $h'_{\mathrm{piop}}$.
};

\node[box] (vf3) at (\xC,\ythree) {
\textbf{V3: Accept / reject}\\
Accept iff
$h'_{\mathrm{piop}}=h_{\mathrm{piop}}$
and $v_{\mathrm{grinding}}=0$.
};

\draw[arrow] (vftitle) -- (vf1);
\draw[arrow] (vf1) -- (vf2);
\draw[arrow] (vf2) -- (vf3);

% -------------------------
% Dashed arrows between columns
% -------------------------
% -------------------------
% Dashed arrows between columns
% -------------------------

% KeyGen -> Sign
\draw[darrow]
    (kg3.east) -- ++(0.15,0)
    coordinate (kgsgbend)
    |- (sg1.west);

\node[
    rotate=90,
    fill=white,
    inner sep=1pt
]
at ($(kgsgbend)!0.5!(kgsgbend |- sg1.west)$)
{$pk,\ sk$};

% Sign -> Verify
\draw[darrow]
    (sg3.east) -- ++(0.15,0)
    coordinate (sgvfbend)
    |- (vf1.west);

\node[
    rotate=90,
    fill=white,
    inner sep=1pt
]
at ($(sgvfbend)!0.5!(sgvfbend |- vf1.west)$)
{$pk,\ msg,\ \sigma$};

\end{tikzpicture}%
}
\caption{Overview of Mirath's \texttt{KeyGen}, \texttt{Sign}, and \texttt{Verify} operations.}
\label{fig:mirath_overview}
\end{figure}
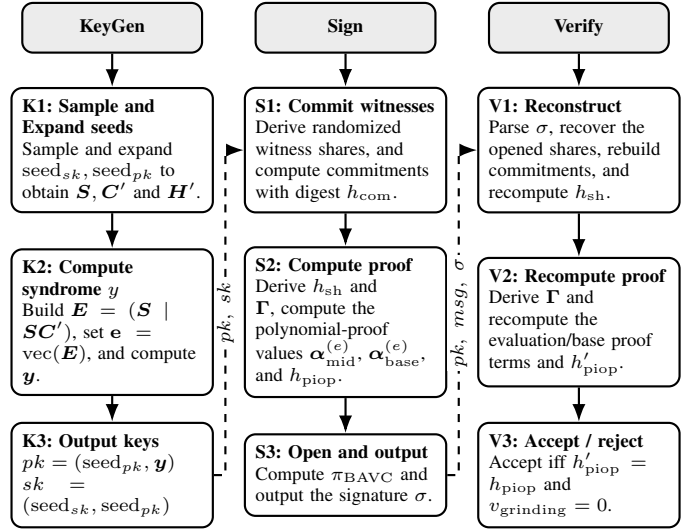

A digital signature algorithm (DSA) consists of the operations \texttt{KeyGen}, \texttt{Sign}, and \texttt{Verify}. 
MPC-in-the-Head (MPCitH) signatures locally simulate an MPC protocol
among virtual parties and commit to their views. 
Threshold Computation-in-the-Head (TCitH) replaces additive secret sharing with packed Shamir secret sharing~\cite{franklin1992communication} to reduce the opening data. 
Mirath is a TCitH-based scheme. 
Although it did not advance to the third round of the NIST process due to similarities with other candidates, its security claims remain unaffected. Fig.~\ref{fig:mirath_overview} summarizes its computation flow.

Mirath is based on the MinRank Syndrome Problem~\cite{bidoux2024dual}.
Let $q=p^s$ be a prime power, with $p$ prime and $s\in\mathbb{N}_{>0}$, and let
$m,n,k,r,\mu,\rho,\tau,N,\lambda\in\mathbb{N}_{>0}$ denote the MinRank dimensions, extension degree, number of parallel polynomial checks, protocol repetitions, virtual parties per repetition, and seed security parameter, respectively. 
Given $\boldsymbol{H}=[\boldsymbol{I}_{mn-k}\parallel\boldsymbol{H}'] \in\mathbb{F}_q^{(mn-k)\times mn}$,
$\boldsymbol{H}'\in\mathbb{F}_q^{(mn-k)\times k}$, and $\boldsymbol{y}\in\mathbb{F}_q^{mn-k}$, the witness is $\boldsymbol{E}\in\mathbb{F}_q^{m\times n}$ satisfying $\boldsymbol{H}\operatorname{vec}(\boldsymbol{E})=\boldsymbol{y}$ and $\operatorname{rank}(\boldsymbol{E})\leq r$. 
It is represented as $\boldsymbol{E}=\boldsymbol{S} [\boldsymbol{I}_r\parallel\boldsymbol{C}']$, with $\boldsymbol{S}\in\mathbb{F}_q^{m\times r}$ and $\boldsymbol{C}'\in\mathbb{F}_q^{r\times(n-r)}$.

During \texttt{KeyGen}, $\mathrm{seed}_{sk},\mathrm{seed}_{pk}\in\{0,1\}^{\lambda}$ are expanded to derive $(\boldsymbol{S},\boldsymbol{C}')$ and $\boldsymbol{H}'$, from which $\boldsymbol{y}$ is computed. 
During \texttt{Sign}, for $e\in[1,\tau]$ and $i\in[1,N]$, a Goldreich–Goldwasser–Micali (GGM) tree derives leaf seeds $\mathrm{seed}^{(e)}_i\in\{0,1\}^{\lambda}$. Each seed is committed as $\mathrm{com}^{(e)}_i$, with all commitments bound by the digest $h_{\mathrm{com}}$, and expanded into $\boldsymbol{S}^{(e)}_{\mathrm{rnd},i}\in\mathbb{F}_q^{m\times r}$, $\boldsymbol{C}'^{(e)}_{\mathrm{rnd},i}\in \mathbb{F}_q^{r\times(n-r)}$, and $\boldsymbol{v}^{(e)}_{\mathrm{rnd},i}\in \mathbb{F}_{q^\mu}^{\rho\times1}$. Using $\boldsymbol{\Gamma}\in \mathbb{F}_{q^\mu}^{\rho\times(mn-k)}$, these values are combined into the polynomial-proof values $\boldsymbol{\alpha}^{(e)}_{\mathrm{mid}}, \boldsymbol{\alpha}^{(e)}_{\mathrm{base}} \in\mathbb{F}_{q^\mu}^{\rho\times1}$.
The BAVC opening $\pi_{\mathrm{BAVC}}$ reveals sufficient seed-tree information to reconstruct all non-challenged parties while keeping the challenged leaf seeds hidden. \texttt{Verify} reconstructs these parties and recomputes the corresponding commitments, shares, and proof values.

At the implementation level, the block cipher Advanced Encryption Standard (AES)~\cite{nist2023aes} is used for GGM-tree expansion, pseudorandom-share generation, and, in this work, leaf-seed commitment generation. 
Federal Information Processing Standard (FIPS) 202~\cite{nist2015fips202}, instantiated with the Secure Hash Algorithm 3 SHA3-256 hash function and the SHAKE128 extendable-output function (XOF), is used for matrix and seed expansion, transcript hashing and challenge generation.

\section{Design Objectives}
\label{sec:design_obj}

Throughout this work, we consider the \texttt{Mirath-1a-fast} parameter set and use Mirath's reference \textsc{C} implementation (denoted as $\mathcal{P}_{\text{Ref}}$ in the following) using test vectors with $33$-byte message size as baseline.
Our hardware platform is based on our proprietary RISC-V core that implements the RV64I with the \texttt{M}, \texttt{C}, \texttt{Zba}, \texttt{Zbb}, \texttt{Zicsr}, and \texttt{Zifencei} extensions and uses a three-stage pipeline, an instruction-prefetch depth of four, and support for four outstanding read requests.

The $\mathcal{P}_{\text{Ref}}$ results in Table~\ref{tab:algorithmic_optimization_results} reveal two main limitations. First, \texttt{Sign} and \texttt{Verify} require substantial data memory, restricting their applicability to resource-constrained embedded systems. Second, their long execution times result in a prohibitive latency for general applications. Throughout this work, latency denotes the computation time of one cryptographic operation. These observations define the two main design objectives of this work: (i) to reduce the memory requirements of Mirath through algorithmic optimization, and (ii) to reduce the resulting execution latency by accelerating the dominant computational bottlenecks. We therefore adopt a cross-layer optimization strategy consisting of an algorithmic layer for memory reduction, followed by an implementation layer based on HW/SW co-design for hardware acceleration.

\begin{table}[t]
    \centering
    \caption{Latency and memory comparison of the reference and
    algorithmically optimized implementations at 500~MHz.}
    \label{tab:algorithmic_optimization_results}

    \footnotesize
    \renewcommand{\arraystretch}{1.15}

    \begin{tabular*}{\columnwidth}{
        @{\extracolsep{\fill}}
        l
        c
        c
        c
        c
        c
        @{}
    }
        \toprule

        \textbf{Function} &
        \textbf{Impl.} &
        \makecell{\textbf{Cycles}\\\textbf{[M]}} &
        \makecell{\textbf{Latency}\\\textbf{[s]}} &
        \makecell{\textbf{Data}\\\textbf{[kB]}} &
        \makecell{\textbf{Code}\\\textbf{[kB]}} \\
        \midrule

        \multirow{2}{*}{\texttt{Sign}}
        & $\mathcal{P}_{\text{Ref}}$
        & 14,471.64
        & 28.94
        & 377
        & 28 \\

        & $\mathcal{P}_0$ (Alg.\,Opt.)
        & 54,019.66
        & 108.04
        & 35
        & 27 \\

        \midrule

        \multirow{2}{*}{\texttt{Verify}}
        & $\mathcal{P}_{\text{Ref}}$
        & 21,476.88
        & 42.95
        & 380
        & 27 \\

        & $\mathcal{P}_0$ (Alg.\,Opt.)
        & 66,321.67
        & 132.64
        & 32
        & 25 \\

        \bottomrule
    \end{tabular*}
    
\end{table}

\section{Algorithmic Optimization}

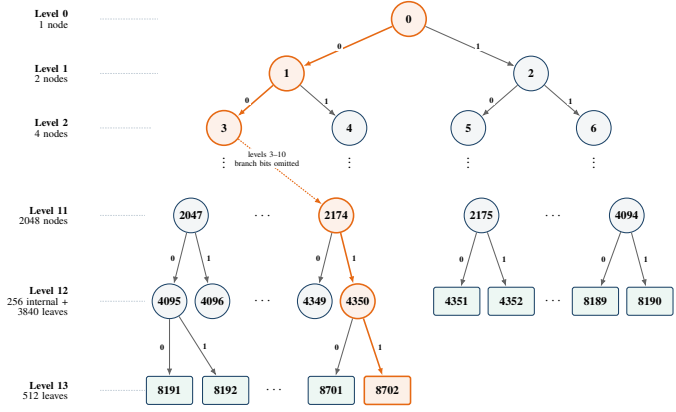
\begin{figure}[t]
    \centering
    \resizebox{\columnwidth}{!}{%
    \begin{tikzpicture}[
        font=\sffamily,
        internal/.style={
            circle, draw=seedblue, fill=seedblue!5,
            minimum size=10.5mm, inner sep=1pt,
            align=center, line width=0.9pt,
            font=\bfseries, text=black
        },
        leaf/.style={
            rounded corners=2pt, draw=seedblue, fill=leafgreen,
            minimum width=14mm, minimum height=8.5mm,
            align=center, line width=0.9pt,
            font=\bfseries, text=black
        },
        hinternal/.style={
            internal, draw=pathorange, fill=pathorange!9,
            line width=1.35pt, text=black
        },
        hleaf/.style={
            leaf, draw=pathorange, fill=pathorange!10,
            line width=1.35pt, text=black
        },
        edge/.style={
            -{Latex[length=2.2mm]}, draw=gray!78!black,
            line width=0.9pt
        },
        hedge/.style={
            -{Latex[length=2.2mm]}, draw=pathorange,
            line width=1.35pt
        },
        omittededge/.style={
            -{Latex[length=2.2mm]}, densely dotted,
            line width=1.0pt
        },
        leveltext/.style={
            anchor=east, align=right, text=black,
            font=\bfseries\small
        },
        note/.style={
            rounded corners=2pt, draw=seedblue, fill=seedblue!3,
            align=left, inner sep=5pt, font=\small,
            text=black
        },
        branchbit/.style={
            midway,
            fill=white,
            inner sep=1.2pt,
            font=\bfseries\scriptsize,
            text=black
        }
    ]

    % Level labels
    \node[leveltext] at (-10.2, 0.0)
        {Level 0\\[-1pt]\normalfont 1 node};
    \node[leveltext] at (-10.2,-1.65)
        {Level 1\\[-1pt]\normalfont 2 nodes};
    \node[leveltext] at (-10.2,-3.30)
        {Level 2\\[-1pt]\normalfont 4 nodes};
    \node[leveltext] at (-10.2,-5.95)
        {Level 11\\[-1pt]\normalfont 2048 nodes};
    \node[leveltext] at (-10.2,-8.55)
        {Level 12\\[-1pt]\normalfont 256 internal +\\[-1pt]
         \normalfont 3840 leaves};
    \node[leveltext] at (-10.2,-11.25)
        {Level 13\\[-1pt]\normalfont 512 leaves};

    \foreach \y/\xend in {
        0/-7.6,
        -1.65/-7.6,
        -3.30/-7.8,
        -5.95/-8.0,
        -8.55/-8.0,
        -11.25/-8.0}
      \draw[seedblue!55, densely dotted]
        (-9.35,\y) -- (\xend,\y);

    % ------------------------------------------------------------
    % Top of the tree
    % ------------------------------------------------------------
    \node[hinternal] (n0) at (0,0) {0};

    \node[hinternal] (n1) at (-3.7,-1.65) {1};
    \node[internal]  (n2) at ( 3.7,-1.65) {2};

    \node[hinternal] (n3) at (-5.6,-3.30) {3};
    \node[internal]  (n4) at (-1.8,-3.30) {4};
    \node[internal]  (n5) at ( 1.8,-3.30) {5};
    \node[internal]  (n6) at ( 5.6,-3.30) {6};

    % Branch bit: left = 0, right = 1
    \draw[hedge] (n0) --
        node[branchbit, xshift=-2.5mm] {0}
        (n1);

    \draw[edge] (n0) --
        node[branchbit, xshift= 2.5mm] {1}
        (n2);

    \draw[hedge] (n1) --
        node[branchbit, xshift=-2.5mm] {0}
        (n3);

    \draw[edge] (n1) --
        node[branchbit, xshift= 2.5mm] {1}
        (n4);

    \draw[edge] (n2) --
        node[branchbit, xshift=-2.5mm] {0}
        (n5);

    \draw[edge] (n2) --
        node[branchbit, xshift= 2.5mm] {1}
        (n6);

    \node[text=black, font=\large] at (-5.6,-4.25) {$\vdots$};
    \node[text=black, font=\large] at (-1.8,-4.25) {$\vdots$};
    \node[text=black, font=\large] at ( 1.8,-4.25) {$\vdots$};
    \node[text=black, font=\large] at ( 5.6,-4.25) {$\vdots$};

    % ------------------------------------------------------------
    % Level 11 boundary
    % ------------------------------------------------------------
    \node[internal]  (p2047) at (-6.6,-5.95) {2047};
    \node[hinternal] (p2174) at (-2.2,-5.95) {2174};
    \node[internal]  (p2175) at ( 2.2,-5.95) {2175};
    \node[internal]  (p4094) at ( 6.6,-5.95) {4094};

    \draw[hedge,omittededge,draw=pathorange]
        (n3) --
        node[
            midway,
            fill=white,
            inner sep=1.5pt,
            text=black,
            font=\scriptsize,
            align=center,
            xshift=-4mm,
            yshift=4mm
        ] {levels 3--10\\branch bits omitted}
        (p2174);

    \node[text=black, font=\large] at (-4.4,-5.95) {$\cdots$};
    \node[text=black, font=\large] at ( 4.4,-5.95) {$\cdots$};

    % ------------------------------------------------------------
    % Level 12
    % ------------------------------------------------------------
    \node[internal]  (i4095) at (-7.25,-8.55) {4095};
    \node[internal]  (i4096) at (-5.95,-8.55) {4096};

    \node[internal]  (i4349) at (-2.85,-8.55) {4349};
    \node[hinternal] (i4350) at (-1.55,-8.55) {4350};

    \draw[edge] (p2047) --
        node[branchbit, xshift=-2.2mm] {0}
        (i4095);

    \draw[edge] (p2047) --
        node[branchbit, xshift= 2.2mm] {1}
        (i4096);

    \draw[edge] (p2174) --
        node[branchbit, xshift=-2.2mm] {0}
        (i4349);

    \draw[hedge] (p2174) --
        node[branchbit, xshift= 2.2mm] {1}
        (i4350);

    \node[text=black, font=\large] at (-4.4,-8.55) {$\cdots$};

    \node[leaf] (l4351) at (1.45,-8.55) {4351};
    \node[leaf] (l4352) at (3.10,-8.55) {4352};

    \node[leaf] (l8189) at (5.65,-8.55) {8189};
    \node[leaf] (l8190) at (7.30,-8.55) {8190};

    \draw[edge] (p2175) --
        node[branchbit, xshift=-2.2mm] {0}
        (l4351);

    \draw[edge] (p2175) --
        node[branchbit, xshift= 2.2mm] {1}
        (l4352);

    \draw[edge] (p4094) --
        node[branchbit, xshift=-2.2mm] {0}
        (l8189);

    \draw[edge] (p4094) --
        node[branchbit, xshift= 2.2mm] {1}
        (l8190);

    \node[text=black, font=\large] at (4.40,-8.55) {$\cdots$};

    % ------------------------------------------------------------
    % Level 13
    % ------------------------------------------------------------
    \node[leaf] (l8191) at (-7.25,-11.25) {8191};
    \node[leaf] (l8192) at (-5.55,-11.25) {8192};

    \draw[edge] (i4095) --
        node[branchbit, xshift=-2.2mm] {0}
        (l8191);

    \draw[edge] (i4095) --
        node[branchbit, xshift= 2.2mm] {1}
        (l8192);

    \node[leaf]  (l8701) at (-2.35,-11.25) {8701};
    \node[hleaf] (l8702) at (-0.65,-11.25) {8702};

    \draw[edge] (i4350) --
        node[branchbit, xshift=-2.2mm] {0}
        (l8701);

    \draw[hedge] (i4350) --
        node[branchbit, xshift= 2.2mm] {1}
        (l8702);

    \node[text=black, font=\large] at (-4.05,-11.25) {$\cdots$};

    \end{tikzpicture}%
    }
    \caption{Compact layout of the Mirath-1a-fast GGM seed tree. The 4352 leaves do not form a perfect binary tree and therefore span two depths. The orange nodes and arrows show one exact root-to-leaf path, ending at node 8702. At each internal node, the 16-byte parent seed is used as the AES-128 key. Each branch uses a distinct 16-byte plaintext block derived from the salt, node index, and branch bit, and the 16-byte ciphertext becomes the child seed.}
    \label{fig:mirath_seed_tree}
\end{figure}

The main memory bottleneck is the pre-computation of round inputs. During \texttt{Sign}, the complete GGM tree and leaf commitments each require approximately $139$\,kB, while duplicating the $4352$ leaf seeds for $h_{\mathrm{com}}$ adds $70$\,kB to the $377$\,kB data-memory footprint. 
We instead derive leaves and commitments on demand from the root along its corresponding path (Fig.~\ref{fig:mirath_seed_tree}), immediately use and discard them, and incrementally absorb commitments into $h_{\mathrm{com}}$. \texttt{Verify} applies the same strategy to visible leaves reconstructed from the sibling seeds in $\pi_{\mathrm{BAVC}}$.

This introduces a memory--computation trade-off: the optimization replaces the full tree and commitment arrays with one $224$\,B path and one $32$\,B commitment, reducing \texttt{Sign}/\texttt{Verify} data memory by approximately $90\,\%$. Repeated tree traversals, however, increase AES evaluations and latency by up to $273\,\%$, motivating hardware acceleration. The comparison between $\mathcal{P}_{\text{Ref}}$ and the algorithmically optimized version, denoted as $\mathcal{P}_0$, is shown in~Table~\ref{tab:algorithmic_optimization_results}.

\section{HW/SW Co-Design}

\begin{table}[t]
    \centering
    \caption{Profiling results for \texttt{KeyGen}, \texttt{Sign}, and \texttt{Verify}. Values are reported as percentages of the total execution time.}
    \label{tab:profiling_results}

    \renewcommand{\arraystretch}{1.15}

    \begin{tabular*}{\columnwidth}{@{\extracolsep{\fill}}lrrr@{}}
        \toprule
        \textbf{Computational Kernel}
        & \textbf{KeyGen (\%)}
        & \textbf{Sign (\%)}
        & \textbf{Verify (\%)} \\
        \midrule

        AES
        & 0.00
        & 98.65
        & 99.36 \\

        FIPS~202
        & 1.59
        & 0.01
        & 0.01 \\

        GF Arithmetic
        & 98.35
        & 1.32
        & 0.62 \\

        Others
        & 0.06
        & 0.03
        & 0.02 \\

        \bottomrule
    \end{tabular*}
\end{table}

\begin{figure}
	\centering
		\includegraphics[width=0.3\textwidth]{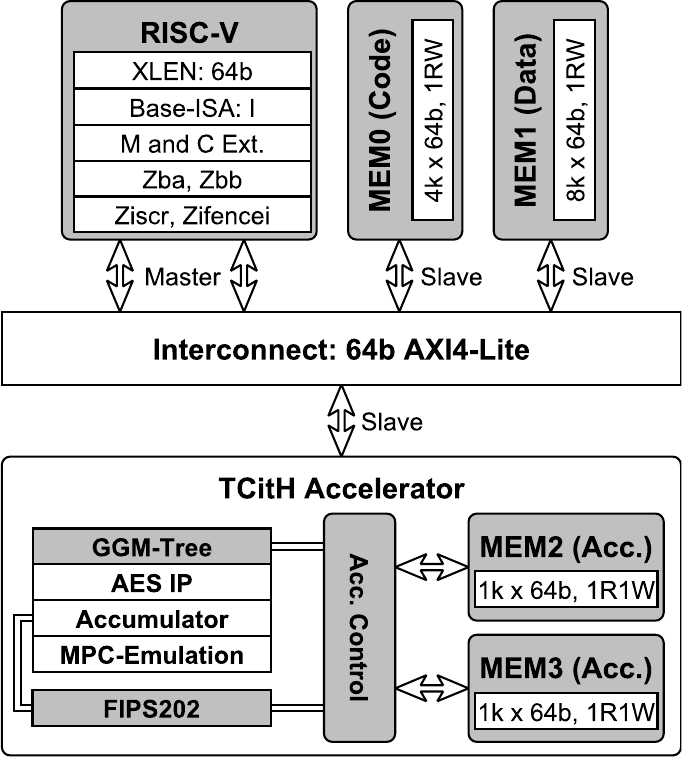}
		\caption{RISC-V architecture and TCitH accelerator unit after memory optimization for the $\mathcal{P}_{0}$--$\mathcal{P}_{12}$ configurations.}
	\label{fig:block_diagram}
\end{figure}

We profiled the optimized software implementation to identify computational bottlenecks and guide HW/SW partitioning. As shown in Table~\ref{tab:profiling_results}, \texttt{KeyGen} is dominated by GF arithmetic, whereas \texttt{Sign} and \texttt{Verify} are dominated by AES operations associated with GGM-tree traversal and commitment generation. Based on these results, we developed a suite of hardware accelerators. The overall system architecture, comprising the processor and the TCitH accelerator, is illustrated in Fig. \ref{fig:block_diagram}. The processor communicates with the accelerator through an AXI4-Lite interface. Local storage is provided by two $8$\,kB memories to hold operands, intermediate values, and results.

Software controls data transfers, accelerator configuration, execution, and result retrieval. Intermediate values are retained locally and, where possible, transferred directly between accelerators to reduce data movement. 
A memory-mapped interface is used instead of custom RISC-V instructions because the targeted kernels involve multiple wide operands. 
The accelerator modules implement progressively coarser offload granularities. While the standalone AES module processes one block per software request and the GGM walker derives a single requested tree node, the Accumulator evaluates all parties of one protocol repetition, and the MPC-Emulation module executes complete algebraic kernels.

\subsection{AES IP}
The system integrates an open-source AES IP core \cite{swann2021reconfigurable}. Although the core supports AES-128/256 encryption and decryption, only AES-128 encryption is utilized for the targeted parameter set. The remaining modes are preserved to maintain versatility for broader application contexts. The AES core is unused during \texttt{KeyGen}. During \texttt{Sign} and \texttt{Verify}, the core accelerates GGM seed derivation, leaf-seed commitments, and pseudorandom-share generation.

\subsection{GGM Tree Accelerator}
\label{subsec:ggm_tree}
In the GGM Tree module, the AES IP is scheduled internally, eliminating the need for one processor command per AES block. The accelerator receives a root seed, salt, and node or leaf index, determines the corresponding left/right directions of the path, and applies the AES-based child derivation function only along the selected path. It neither generates both children nor materializes the complete tree. 

The local memories retain the preceding tree path, allowing consecutive traversals with a common prefix to resume from the first differing level.

This accelerator is not used during \texttt{KeyGen}. During \texttt{Sign}, it performs the root-to-leaf traversals and derives the sibling-node and hidden-party seeds required for $\pi_{\mathrm{BAVC}}$. During \texttt{Verify}, reconstruction starts from the sibling-path seeds disclosed in the signature rather than from the original root. This mode is integrated only when the Accumulator is used. Without it, software traverses from the disclosed seeds by invoking the AES IP for the required tree edges.

\subsection{FIPS~202 Accelerator}

The FIPS~202 hardware module processes aligned 64-bit portions of the absorb and squeeze operations and executes the \textsc{Keccak} permutations. Therefore, software retains control of the incremental context, unaligned input and output fragments, padding, byte extraction, and the protocol-specific construction of each hash input. Additionally, when combined with the Accumulator, 32-byte party commitments are transferred directly to the \textsc{Keccak} input through a 64-bit ready/valid interface. This avoids returning every commitment to the processor and subsequently writing it back to the hash accelerator.

The FIPS~202 accelerator is used for three classes of operations: expansion of public and secret matrices; computation of commitment and transcript digests; and derivation of the protocol challenges.

\subsection{Accumulator}
The Accumulator is always combined with the GGM Tree module. It is not used during \texttt{KeyGen}. During \texttt{Sign}, software loads the salt, root seed, and secret matrices and issues one command per repetition $e$. The accelerator derives the party seeds and commitments, expands the shares, and computes the accumulations, party-weighted bases, and auxiliary values.

AES seed generation and share processing are overlapped such that one AES block can be accumulated while the next is generated. Extension-field arithmetic uses eight parallel byte-wide $\mathbb{F}_{2^8}$ multipliers, processing one 64-bit word per arithmetic step. Base-field elements in $\mathbb{F}_{16}$ use the same embedded representation as $\mathcal{P}_{\text{Ref}}$. The resulting auxiliary and base values remain in the local memories and can be consumed directly by the MPC-Emulation accelerator when it is also enabled.

During \texttt{Verify}, software parses $\pi_{\mathrm{BAVC}}$ into a path table containing the disclosed sibling seeds and the leaf ranges covered by their subtrees. For each visible party, the accelerator selects the corresponding entry and completes the remaining GGM traversal from the disclosed seed. It then reconstructs the visible parties, expands their shares, and regenerates their commitments. For the hidden parties, the auxiliary values and commitment are obtained from the signature. When direct commitment streaming is enabled, the hidden commitments are inserted at the corresponding position in the stream to the FIPS~202 accelerator.

\subsection{MPC-Emulation Accelerator}
The MPC-Emulation accelerator performs the algebraic proof computations after share accumulation. It is implemented as a multi-phase state machine and reuses the same eight $\mathbb{F}_{2^8}$ multiplier lanes as the Accumulator. An early-exit mode reuses the matrix-product datapath to compute $y$. This mode is used for syndrome computation during \texttt{KeyGen} and in \texttt{Sign} during a reconstruction of $y$ as an implementation-specific prerequisite for $h_{\mathrm{piop}}$. The full mode computes the polynomial-proof values during \texttt{Sign} and \texttt{Verify}. It is not used during \texttt{KeyGen}.

\subsection{Partitionings}
The previously described accelerator functionalities can be selectively implemented in software or hardware, resulting in multiple HW/SW partitioning configurations. This work considers $14$ distinct partitions, shown in~Table~\ref{tab:hwsw_partitionings}. Since \texttt{KeyGen} was not algorithmically optimized, $\mathcal{P}_{\text{Ref}}$ and $\mathcal{P}_0$ are the same, and only $\mathcal{P}_2$, $\mathcal{P}_3$, and $\mathcal{P}_5$ are applicable. For \texttt{Sign}, all partitions are supported, whereas for \texttt{Verify}, all partitions except $\mathcal{P}_4$ are applicable, as discussed in Section~\ref{subsec:ggm_tree}. Subsequent references to $\mathcal{P}_{12}$ also include \texttt{KeyGen}, which uses only its supported accelerators.

\section{Results}
The ASIC results are based on the GlobalFoundries 22\,nm FD-SOI technology. Timing is evaluated under worst-case PVT conditions at $125\,^\circ$C and $0.72$\,V, whereas power is evaluated under nominal conditions at $25\,^\circ$C and $0.8$\,V. The design flow employed Synopsys DesignCompiler, IC-Compiler, and the INVECAS Memory Compiler, with power calculated via back-annotated wiring data.

\subsection{Impact of the different partitions on the overall runtime}

The impact of the HW/SW partitions on execution time and memory requirements is summarized in Table~\ref{tab:mirath_hw_modules} for \texttt{Sign} and \texttt{Verify}. For KeyGen, latency decreases from 3.07 ms in $\mathcal{P}_0$ to 2.67 ms, 0.60 ms, and 0.20 ms in $\mathcal{P}_2$, $\mathcal{P}_3$, and $\mathcal{P}_5$, respectively. 
All partitions were validated against $\mathcal{P}_{\text{Ref}}$. The results show that the effectiveness of each accelerator strongly depends on the targeted operation and on the remaining software bottlenecks.

\begin{table}[t]
    \centering
    \caption{HW/SW partitions and their corresponding hardware accelerators.}
    \label{tab:hwsw_partitionings}
    \footnotesize

    \begin{tabular*}{\columnwidth}
        {@{\extracolsep{\fill}}cl@{}}
        \toprule
        \textbf{Partition} & \textbf{Hardware Accelerators} \\
        \midrule

        $\mathcal{P}_{\text{Ref}}$ & None (Reference Implementation) \\
        $\mathcal{P}_0$ & None (Algorithmically Optimized Software) \\
        $\mathcal{P}_1$ & AES IP \\
        $\mathcal{P}_2$ & FIPS~202 \\
        $\mathcal{P}_3$ & MPC-Emulation \\
        $\mathcal{P}_4$ & AES IP, GGM Tree \\
        $\mathcal{P}_5$ & FIPS~202, MPC-Emulation\\
        $\mathcal{P}_6$ & AES IP, GGM Tree, FIPS~202 \\
        $\mathcal{P}_7$ & AES IP, GGM Tree, Accumulator \\
        $\mathcal{P}_8$ & AES IP, GGM Tree, MPC-Emulation \\
        $\mathcal{P}_9$ & AES IP, GGM Tree, FIPS~202, Accumulator \\
        $\mathcal{P}_{10}$ & AES IP, GGM Tree, FIPS~202, MPC-Emulation \\
        $\mathcal{P}_{11}$ & AES IP, GGM Tree, Accumulator, MPC-Emulation \\
        $\mathcal{P}_{12}$ & AES IP, GGM Tree, FIPS~202, Accumulator, MPC-Emulation \\
        \bottomrule
    \end{tabular*}
\end{table}

\begin{table}[t]
    \centering
    \caption{ASIC Signing and verification latency and number of clock cycles of Mirath for the considered HW/SW partitions.}
    \label{tab:mirath_hw_modules}

    \renewcommand{\arraystretch}{0.95}
    \setlength{\aboverulesep}{0.3ex}
    \setlength{\belowrulesep}{0.3ex}

    \begin{tabular*}{\columnwidth}{@{\extracolsep{\fill}}lrrrr@{}}
        \toprule
        &
        \multicolumn{2}{c}{\textbf{Sign}} &
        \multicolumn{2}{c}{\textbf{Verify}} \\
        \cmidrule(lr){2-3}
        \cmidrule(lr){4-5}

        \textbf{Impl.}
        & \textbf{\shortstack{Cycles\\(M)}}
        & \textbf{\shortstack{Latency\\(ms)}}
        & \textbf{\shortstack{Cycles\\(M)}}
        & \textbf{\shortstack{Latency\\(ms)}} \\
        \midrule

        $\mathcal{P}_{\text{Ref}}$
        & 14,471.64 & 28,943.28
        & 21,476.88 & 42,953.76 \\

        $\mathcal{P}_0$
        & 54,019.66 & 108,039.32
        & 66,321.67 & 132,643.34 \\

        \midrule
        \multicolumn{5}{c}{\textbf{One Accelerator}} \\

        $\mathcal{P}_1$
        & 212.56 & 425.12
        & 200.35 & 400.69 \\

        $\mathcal{P}_2$
        & 54,017.33 & 108,034.66
        & 63,824.19 & 127,648.37 \\

        $\mathcal{P}_3$
        & 51,912.17 & 103,824.33
        & 66,306.51 & 132,613.01 \\

        \midrule
        \multicolumn{5}{c}{\textbf{Two Accelerators}} \\

        $\mathcal{P}_4$
        & 165.58 & 331.16
        & \multicolumn{2}{c}{Same as $\mathcal{P}_1$$^\ddagger$}\\

        $\mathcal{P}_5$
        & 51,909.78 & 103,819.56
        & 63,750.99 & 127,501.97 \\

        \midrule
        \multicolumn{5}{c}{\textbf{Three Accelerators}} \\

        $\mathcal{P}_6$
        & 163.20 & 326.39
        & 194.35 & 388.70 \\

        $\mathcal{P}_7$
        & 79.59 & 159.18
        & 44.10 & 88.20 \\

        $\mathcal{P}_8$
        & 86.39 & 172.77
        & 155.18 & 310.36 \\

        \midrule
        \multicolumn{5}{c}{\textbf{Four Accelerators}} \\

        $\mathcal{P}_9$
        & 72.44 & 144.87
        & 34.70 & 69.40 \\

        $\mathcal{P}_{10}$
        & 84.05 & 168.10
        & 148.96 & 297.92 \\

        $\mathcal{P}_{11}$
        & 8.65 & 17.30
        & 10.80 & 21.60 \\

        \midrule
        \multicolumn{5}{c}{\textbf{All Accelerators}} \\

        $\mathcal{P}_{12}$
        & 1.50 & 2.99
        & 1.39 & 2.78 \\

        \bottomrule

    \multicolumn{5}{l}{$^\ddagger$ For Verify, the GGM Tree accelerator is inactive without
    } \\
    \multicolumn{5}{l}{the Accumulator, so P4 reduces to P1.} \\
    \end{tabular*}
\end{table}
\begin{figure}
	\centering
		\includegraphics[width=0.36\textwidth]{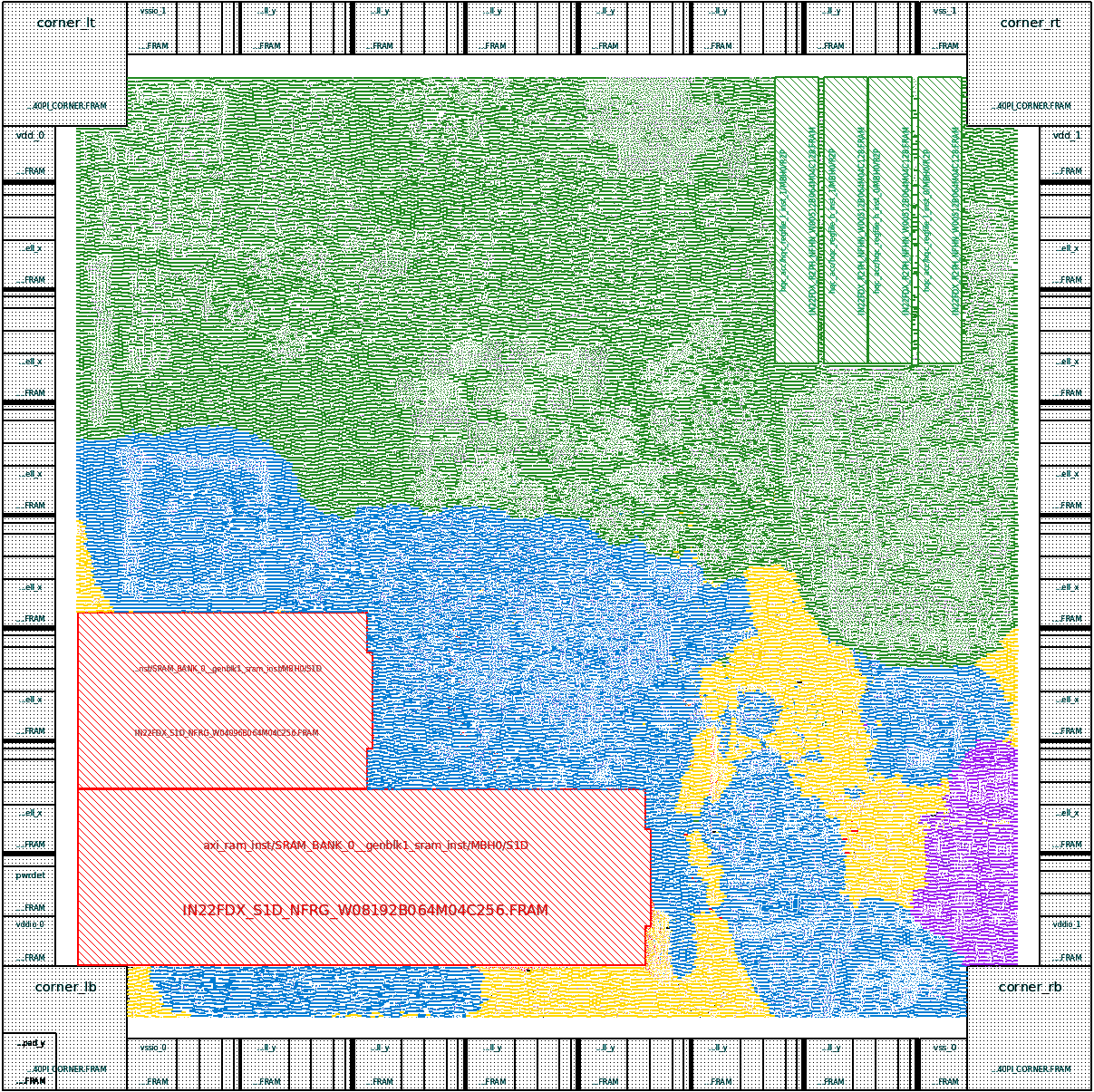}
		\caption{Layout of the fully accelerated $\mathcal{P}_{12}$ ASIC implementation. TCitH accelerator and local memories in green, RISC-V in blue, memory interconnect in yellow, main memories in red, and others in purple.}
	\label{fig:asic_opt}
\end{figure}
\begin{table}[t]
    \centering
    \caption{ASIC area breakdown.}
    \label{tab:asic_area_breakdown}

    \renewcommand{\arraystretch}{1.1}

    \begin{tabular*}{\columnwidth}{@{\extracolsep{\fill}}lrr@{}}
        \toprule
        & \textbf{Area} ($\mu m^2$) & \textbf{Area ($\%$)} \\
        \midrule

        RISC-V & 10{,}741.25 & 4.57 \\
        Interconnect & 6{,}186.55 & 2.63 \\
        Memories & 158{,}841.05 & 67.57 \\
    
        \quad $\vdash$ Main Memories & 119{,}712.39 & 50.92 \\
        \quad $\vdash$ Accelerator Memories & 39{,}128.66 & 16.64 \\
    
        TCitH Accelerator & 42{,}686.92 & 18.16 \\
        \quad $\vdash$ Control Logic & 2{,}513.97 & 1.07 \\
        \quad $\vdash$ FIPS~202 & 9{,}795.44 & 4.17 \\
        \quad $\vdash$ Combined Modules & 30{,}377.52 & 12.92 \\
    
        \quad\quad $\vdash$ GGM Tree & 8{,}486.47 & 3.61 \\
        \quad\quad $\vdash$ AES IP & 11{,}172.56 & 4.75 \\
        \quad\quad $\vdash$ Accumulator & 7{,}040.92 & 3.00 \\
        \quad\quad $\vdash$ MPC-Emulation & 3{,}677.57 & 1.56 \\
    
        Others & 16{,}635.54 & 7.08 \\
    
        \midrule
        $\mathcal{P}_{12}$ & 235{,}091.33 & 100.00 \\

        \bottomrule
    \end{tabular*}
\end{table}
\begin{table*}[h!]
    \centering
    \caption{Comparison of our work and related HW/SW Co-Design based ASIC implementations of NIST standardized DSAs. All implementations correspond to the smallest parameter set of the respective algorithms. n.r. denotes not reported.}
    \label{tab:comparison_related_work}

    \begin{tabular*}{\textwidth}{@{\extracolsep{\fill}}lcccc*{2}{c}*{2}{c}*{2}{c}@{}}
        \toprule
    
        \multirow{2}{*}{\textbf{Work}} &
        \multirow{2}{*}{\makecell{\textbf{Technology}\\\textbf{[nm]}}} &
        \multicolumn{2}{c}{\textbf{Area}} &
        \multirow{2}{*}{\makecell{\textbf{Freq.}\\\textbf{[MHz]}}} &
        \multicolumn{2}{c}{\textbf{KeyGen}} &
        \multicolumn{2}{c}{\textbf{Sign}} &
        \multicolumn{2}{c}{\textbf{Verify}} \\
    
        \cmidrule(lr){3-4}
        \cmidrule(lr){6-7}
        \cmidrule(lr){8-9}
        \cmidrule(lr){10-11}
    
        & &
        \textbf{[mm$^2$]} &
        \textbf{[kGE]*} &
        &
        \makecell{\textbf{Mcycles}} &
        \makecell{\textbf{Lat.}\\\textbf{[ms]}} &
        \makecell{\textbf{Mcycles}} &
        \makecell{\textbf{Lat.}\\\textbf{[ms]}} &
        \makecell{\textbf{Mcycles}} &
        \makecell{\textbf{Lat.}\\\textbf{[ms]}} \\

        \midrule

        \textbf{This work ($\mathcal{P}_{12}$)}$^{\dagger}$ &
        22 &
        0.24 &
        381.86 &
        500 &
        0.10 & 0.20 &
        1.50 & 2.99 &
        1.39 & 2.78 \\

        \midrule
        \multicolumn{11}{c}{SPHINCS+/SLH-DSA} \\
        \midrule
        
        Karl et al.~\cite{karl2025performance}
        & 22
        & 0.56
        & 403.30
        & 800
        & 1.73
        & 2.16
        & 42.60
        & 53.25
        & 2.46
        & 3.07 \\

        Saarinen~\cite{saarinen2024sloth} &
        45 &
        n.r. &
        73.08 &
        250 &
        0.18 & 0.71 &
        4.90 & 19.62 &
        0.44 & 1.76 \\

        Dolmeta et al.~\cite{dolmeta2026horcrux}$^\circ$ &
        65 &
        n.r. &
        115.80 &
        160 &
        1.91 & 11.91 &
        45.34 & 283.34 &
        2.93 & 18.31 \\

        \midrule
        \multicolumn{11}{c}{Dilithium/ML-DSA} \\
        \midrule

        Karl et al.~\cite{karl2024riscv}$^{\dagger}$ &
        22 &
        0.46 &
        244.00 &
        800 &
        0.59 & 0.74 &
        1.91 & 2.38 &
        0.65 & 0.81 \\

        Carril et al.~\cite{carril2026pqcuark}$^{\diamond}$ &
        22 &
        2.74 &
        13712.00 &
        1200 &
        0.34 & 0.27 &
        0.91 & 0.72 &
        0.36 & 0.29 \\

        Dolmeta et al.~\cite{dolmeta2026horcrux}$^\circ$ &
        65 &
        n.r. &
        115.80 &
        160 &
        0.46 & 2.86 &
        1.54 & 9.61 &
        0.59 & 3.70 \\

        \midrule
        \multicolumn{11}{c}{Falcon} \\
        \midrule

        Dolmeta et al.~\cite{dolmeta2026horcrux}$^\circ$ &
        65 &
        n.r. &
        115.80 &
        160 &
        117.79 & 736.16 &
        48.61 & 303.81 &
        0.26 & 1.61 \\

        \bottomrule
    \multicolumn{11}{l}{*: kGE values follow the memory accounting of the respective works. $\circ$: only accelerator area reported (excludes processor).} \\
    \multicolumn{11}{l}{$\dagger$: post-place-and-route data (post-synthesis for the others). $\diamond$: reported area does not include the full memory.} \\
    \end{tabular*}
    
\end{table*}

For \texttt{Sign} and \texttt{Verify}, accelerating FIPS~202 or MPC-Emulation alone provides negligible benefit, leaving the system still $197$--$273\,\%$ slower than $\mathcal{P}_{\text{Ref}}$. In contrast, introducing the standalone AES IP with $\mathcal{P}_1$ reduces the latency of $\mathcal{P}_0$ by $99.61\,\%$ and $99.70\,\%$ for
\texttt{Sign} and \texttt{Verify}, respectively, confirming AES as the dominant bottleneck. This improvement is achieved at low hardware cost, with the AES IP accounting for only $4.75\,\%$ of the total $\mathcal{P}_{12}$ cell area.

Adding the GGM Tree accelerator in $\mathcal{P}_4$ further reduces signing latency by approximately $22.1\,\%$ relative to $\mathcal{P}_1$. Although its marginal gain is smaller than that of the standalone AES IP, it removes another major component of the seed-tree computation and enables larger improvements from the Accumulator and MPC-Emulation accelerators.

Additionally, the results show that the FIPS~202 module becomes relevant only after the dominant bottlenecks have been addressed. Although it is the second-largest accelerator after AES in area, its impact is minor in intermediate partitions. However, once the other dominant workloads are accelerated, FIPS~202 becomes the primary remaining bottleneck. Adding it to $\mathcal{P}_{11}$ to obtain the fully accelerated $\mathcal{P}_{12}$ reduces signing and verification cycles by a further $82.7\,\%$ and $87.1\,\%$, respectively. A similar dependency is observed for \texttt{KeyGen}.

\subsection{Area, Power, and Energy}
\begin{table}[h!]
    \centering
    \caption{Power and energy consumption of the proposed Mirath implementations and SoA comparison. n.r. denotes not reported.}
    \label{tab:mirath_power_energy}

    \renewcommand{\arraystretch}{1.15}
    \setlength{\tabcolsep}{3pt}

    \begin{tabular*}{\columnwidth}{@{\extracolsep{\fill}}lcccccc@{}}
        \toprule

        \multirow{2}{*}{\textbf{Implementation}} &
        \multicolumn{2}{c}{\textbf{KeyGen}} &
        \multicolumn{2}{c}{\textbf{Sign}} &
        \multicolumn{2}{c}{\textbf{Verify}} \\

        \cmidrule(lr){2-3}
        \cmidrule(lr){4-5}
        \cmidrule(lr){6-7}

        &
        \makecell{\textbf{Power}\\\textbf{[mW]}} &
        \makecell{\textbf{Energy}\\\textbf{[$\mu$J]}} &
        \makecell{\textbf{Power}\\\textbf{[mW]}} &
        \makecell{\textbf{Energy}\\\textbf{[$\mu$J]}} &
        \makecell{\textbf{Power}\\\textbf{[mW]}} &
        \makecell{\textbf{Energy}\\\textbf{[$\mu$J]}} \\

        \midrule

        \textbf{This work ($\mathcal{P}_{\text{Ref}}$)} &
        16.50 & 50.66 &
        16.50 & 477,560 &
        17.50 & 751,695 \\

        \textbf{This work ($\mathcal{P}_{12}$)} &
        18.20 & 3.60 &
        26.90 & 80.54 &
        25.60 & 71.12 \\

        Karl et al.~\cite{karl2025performance} &
        n.r. & n.r. &
        n.r. & n.r. &
        48.66 & 9,567 \\

        \bottomrule
    \end{tabular*}
\end{table}

For the implementation, we considered the two representative endpoints of the design space: $\mathcal{P}_{\text{Ref}}$ and the fully accelerated $\mathcal{P}_{12}$ configuration. This enables a direct comparison of the area and energy characteristics of the proposed architecture $\mathcal{P}_{12}$ relative to $\mathcal{P}_{\text{Ref}}$.

To run $\mathcal{P}_{\text{Ref}}$, the memory subsystem requires one 32\,kB SRAM macro for code and six 64\,kB SRAM macros for data. 
The resulting ASIC occupies a core area of $0.61$\,mm$^2$ with a utilization of $55.74\,\%$, an aspect ratio of $2.02$, a total cell area of $0.51$\,mm$^2$, and a maximum frequency of $630$\,MHz.
Given the memory reduction that resulted from the algorithmic optimization, the required SRAM macros were significantly reduced from six to one 64\,kB macro for $\mathcal{P}_{12}$. The ASIC occupies a core area of $0.33$\,mm$^2$ with a utilization of $45.92\,\%$, an aspect ratio of $1.0$, a total cell area of $0.24$\,mm$^2$, and a maximum frequency of $650$\,MHz, limited by the main memory access time (Fig. \ref{fig:asic_opt}). The breakdown of the ASIC area is shown in Table~\ref{tab:asic_area_breakdown}, where only the total cell area is considered. Both designs were implemented and simulated with a $500$\,MHz operating frequency.

Beyond reducing latency, the proposed $\mathcal{P}_{12}$ implementation also substantially reduces energy consumption. As summarized in Table~\ref{tab:mirath_power_energy}, $\mathcal{P}_{12}$ exhibits a slightly higher power consumption than $\mathcal{P}_{\text{Ref}}$, which can be attributed to the additional accelerator logic despite the reduced memory footprint. This increase, however, is outweighed by the substantial reduction in execution time, particularly for \texttt{Sign} and \texttt{Verify}. Consequently, the energy consumption of \texttt{KeyGen}, \texttt{Sign}, and \texttt{Verify} is reduced by $14\times$, $5{,}929\times$, and $10{,}569\times$, respectively.

\section{Comparison to SoA}

Table~\ref{tab:comparison_related_work} compares $\mathcal{P}_{12}$ with SoA ASIC implementations of standardized NIST DSAs and their pre-standardization versions.
Direct area comparisons are limited by differences in technology nodes and implementation stages. The PVT conditions of the compared works are not reported.

As shown, our work $\mathcal{P}_{12}$ achieves $6.6\times$ to $95\times$ lower signing latencies compared with the SPHINCS+/SLH-DSA implementations in \cite{dolmeta2026horcrux, karl2025performance, saarinen2024sloth}. 
Furthermore, our design also reduces the verification latency by approximately $1.1\times$ and $6.6\times$ compared with \cite{karl2025performance} and \cite{dolmeta2026horcrux}, respectively.
The work in \cite{saarinen2024sloth} retains a $1.6\times$ advantage in verification and reports a smaller logic area. However, the area values are not directly comparable, due to differences in technology libraries and implementation stages. The most comparable implementation~\cite{karl2025performance}, which uses the same technology node and accounts for the full system (memory, processor, and accelerator), occupies $2.4\times$ the area of $\mathcal{P}_{12}$.

In summary, with our implementation, Mirath meets the NIST-induced requirement for a non-lattice candidate in the additional DSA standardization process to outperform SLH-DSA with respect to latency when jointly considering the sign and verify operation.
Even though Mirath did not advance to the third round, this is an important finding for TCitH schemes with similar computational structures.

Energy comparisons are limited because \cite{karl2025performance} is the only considered work reporting power and energy for an individual cryptographic operation, and only for \texttt{Verify}. For this operation, $\mathcal{P}_{12}$ requires approximately $1.9\times$ lower power and $135\times$ lower energy.

$\mathcal{P}_{12}$ achieves latencies comparable to those of SoA Dilithium/ML-DSA and Falcon implementations, supporting the feasibility of TCitH-based signatures as viable alternatives from an implementation perspective. It outperforms the Dilithium/ML-DSA implementation in \cite{dolmeta2026horcrux} in all three operations and operates in the same latency range as \cite{karl2024riscv} and \cite{carril2026pqcuark}, while reporting $2\times$ and $11.7\times$ lower area, respectively.
 
\section{Conclusion}
In this work, we investigated the implementation characteristics of Mirath, a TCitH-based post-quantum DSA, for resource-constrained systems. We presented, to the best of our knowledge, the first ASIC implementation of a TCitH-based signature scheme and combined algorithmic optimizations with hardware accelerators to substantially reduce the energy consumption of signing and verification relative to the reference software implementation. Our design achieves lower signing latency than all considered SoA SPHINCS+/SLH-DSA ASIC implementations. At the same technology node, it also requires less area and achieves substantially lower verification energy. Additionally, compared to SoA implementations of lattice-based algorithms Dilithium/ML-DSA and Falcon, we have shown that, from an implementation perspective, a TCitH-based algorithm is a viable alternative in embedded applications in terms of computation time and required area.

\clearpage
\newpage
\bibliographystyle{IEEEtran}
\bibliography{references}

@techreport{mosca2025global,
  author      = {Michele Mosca and Marco Piani},
  title       = {{Quantum Threat Timeline Report 2025}},
  institution = {{Global Risk Institute}},
  address     = {Toronto, ON, Canada},
  year        = {2026},
  month       = mar,
  url         = {https://globalriskinstitute.org/publication/quantum-threat-timeline-report-2025b/}
}

@techreport{NISTFIPS2042024,
  author      = {{National Institute of Standards and Technology}},
  title       = {{Module-Lattice-Based Digital Signature Standard}},
  institution = {{National Institute of Standards and Technology}},
  address     = {Gaithersburg, MD},
  type        = {{Federal Information Processing Standards Publication}},
  number      = {204},
  year        = {2024},
  month       = aug,
  doi         = {10.6028/NIST.FIPS.204},
  url         = {https://doi.org/10.6028/NIST.FIPS.204}
}

@techreport{NISTFIPS2052024,
  author      = {{National Institute of Standards and Technology}},
  title       = {{Stateless Hash-Based Digital Signature Standard}},
  institution = {{National Institute of Standards and Technology}},
  address     = {Gaithersburg, MD},
  type        = {{Federal Information Processing Standards Publication}},
  number      = {205},
  year        = {2024},
  month       = aug,
  doi         = {10.6028/NIST.FIPS.205},
  url         = {https://doi.org/10.6028/NIST.FIPS.205}
}

@techreport{FouqueEtAlFalcon2022,
  author      = {Fouque, Pierre-Alain and Hoffstein, Jeffrey and Kirchner, Paul and Lyubashevsky, Vadim and Pornin, Thomas and Prest, Thomas and Ricosset, Thomas and Seiler, Gregor and Whyte, William and Zhang, Zhenfei},
  title       = {{Falcon: Fast-Fourier Lattice-Based Compact Signatures over NTRU}},
  institution = {{falcon-sign.info}},
  year        = {2022},
  type        = {{Technical report}},
  note        = {{Supporting documentation}},
  url         = {https://falcon-sign.info/falcon.pdf}
}

@misc{NISTAdditionalSignatures2023,
  author       = {{National Institute of Standards and Technology}},
  title        = {{Call for Additional Digital Signature Schemes for the Post-Quantum Cryptography Standardization Process}},
  year         = {2023},
  month        = sep,
  howpublished = {\url{https://csrc.nist.gov/Projects/post-quantum-cryptography/additional-signatures-2023}}
}

@techreport{BenadjilaEtAlMQOM2025,
  author      = {Ryad Benadjila and Charles Bouillaguet and Thibauld Feneuil and Matthieu Rivain},
  title       = {{MQOM: MQ on my Mind: Algorithm Specifications and Supporting Documentation}},
  institution = {{MQOM Team}},
  type        = {{Algorithm specification}},
  year        = {2025},
  month       = sep,
  note        = {{Version 2.1}},
  url         = {https://mqom.org/docs/mqom-v2.1.pdf}
}

@techreport{AragonEtAlRYDE2025,
  author      = {Nicolas Aragon and Magali Bardet and Lo{\"i}c Bidoux and
                 Jes{\'u}s-Javier Chi-Dom{\'i}nguez and Victor Dyseryn and Thibauld Feneuil and
                 Philippe Gaborit and Antoine Joux and Romaric Neveu and Matthieu Rivain and
                 Jean-Pierre Tillich and Adrien Vin{\c{c}}otte},
  title       = {{RYDE Signature Scheme}},
  institution = {{RYDE Team}},
  type        = {{Algorithm specification}},
  year        = {2025},
  month       = sep,
  note        = {{Version 2.1.0}},
  url         = {https://pqc-ryde.org/assets/downloads/ryde_specification_v2.1.0.pdf}
}

@techreport{AdjEtAlMirath2025,
  author      = {Gora Adj and Nicolas Aragon and Stefano Barbero and Magali Bardet and
                 Emanuele Bellini and Lo{\"i}c Bidoux and Jes{\'u}s-Javier Chi-Dom{\'i}nguez and
                 Victor Dyseryn and Andre Esser and Thibauld Feneuil and Philippe Gaborit and
                 Romaric Neveu and Matthieu Rivain and Luis Rivera-Zamarripa and Carlo Sanna and
                 Jean-Pierre Tillich and Javier Verbel and Floyd Zweydinger},
  title       = {{Mirath}},
  institution = {{Mirath Team}},
  type        = {{Algorithm specification}},
  year        = {2025},
  month       = feb,
  note        = {{Version 2.0; merger of MIRA and MiRitH}},
  url         = {https://csrc.nist.gov/csrc/media/Projects/pqc-dig-sig/documents/round-2/spec-files/mirath-spec-round2-web.pdf}
}

@inproceedings{IshaiEtAlZeroKnowledgeMPC2007,
  author    = {Yuval Ishai and Eyal Kushilevitz and Rafail Ostrovsky and Amit Sahai},
  title     = {{Zero-Knowledge from Secure Multiparty Computation}},
  booktitle = {Proceedings of the Thirty-Ninth Annual ACM Symposium on Theory of Computing},
  pages     = {21--30},
  publisher = {ACM},
  year      = {2007},
  month     = jun,
  doi       = {10.1145/1250790.1250794},
  url       = {https://doi.org/10.1145/1250790.1250794}
}

@article{FeneuilRivainTCitH2025,
  author  = {Thibauld Feneuil and Matthieu Rivain},
  title   = {{Threshold Computation in the Head: Improved Framework for Post-Quantum Signatures and Zero-Knowledge Arguments}},
  journal = {Journal of Cryptology},
  volume  = {38},
  number  = {3},
  pages   = {28},
  year    = {2025},
  month   = jul,
  doi     = {10.1007/s00145-025-09543-8},
  url     = {https://doi.org/10.1007/s00145-025-09543-8}
}

@techreport{AlagicEtAlNISTIR8610_2026,
  author      = {Gorjan Alagic and Maxime Bros and Pierre Ciadoux and Quynh Dang and
                 Thinh Hung Dang and John Kelsey and Jacob Lichtinger and Yi-Kai Liu and
                 Carl Miller and Dustin Moody and Rene Peralta and Ray Perlner and
                 Angela Robinson and Hamilton Silberg and Daniel Smith-Tone and Noah Waller},
  title       = {{Status Report on the Second Round of the Additional Digital Signature Schemes for the NIST Post-Quantum Cryptography Standardization Process}},
  institution = {{National Institute of Standards and Technology}},
  address     = {Gaithersburg, MD},
  type        = {{NIST Internal Report}},
  number      = {8610},
  year        = {2026},
  month       = may,
  doi         = {10.6028/NIST.IR.8610},
  url         = {https://doi.org/10.6028/NIST.IR.8610}
}

@inproceedings{bidoux2024dual,
  author    = {Lo{\"i}c Bidoux and Thibauld Feneuil and Philippe Gaborit and Romaric Neveu and Matthieu Rivain},
  title     = {{Dual Support Decomposition in the Head: Shorter Signatures from Rank SD and MinRank}},
  booktitle = {Advances in Cryptology -- ASIACRYPT 2024},
  editor    = {Kai-Min Chung and Yu Sasaki},
  series    = {Lecture Notes in Computer Science},
  volume    = {15485},
  pages     = {38--69},
  publisher = {Springer},
  year      = {2024},
  doi       = {10.1007/978-981-96-0888-1_2},
  url       = {https://doi.org/10.1007/978-981-96-0888-1_2}
}

@inproceedings{schoffel2025hw,
  author    = {Maximilian Sch{\"o}ffel and Hiandra Tomasi and Norbert Wehn},
  title     = {{HW/SW Implementation of MiRitH on Embedded Platforms}},
  booktitle = {2025 IEEE 16th Latin America Symposium on Circuits and Systems (LASCAS)},
  pages     = {1--5},
  publisher = {IEEE},
  year      = {2025},
  doi       = {10.1109/LASCAS64004.2025.10966273},
  url       = {https://doi.org/10.1109/LASCAS64004.2025.10966273}
}

@article{benadjila2026breaking,
  author  = {Ryad Benadjila and Thibauld Feneuil},
  title   = {{Breaking the Myth of MPCitH Inefficiency: Optimizing MQOM for Embedded Platforms}},
  journal = {IACR Transactions on Cryptographic Hardware and Embedded Systems},
  volume  = {2026},
  number  = {3},
  pages   = {279--305},
  year    = {2026},
  month   = jul,
  doi     = {10.46586/tches.v2026.i3.279-305},
  url     = {https://doi.org/10.46586/tches.v2026.i3.279-305}
}

@misc{aranha2025faest,
  author       = {Diego F. Aranha and Johan Degn and Jonathan Eilath and Kent Nielsen and Peter Scholl},
  title        = {{FAEST for Memory-Constrained Devices with Side-Channel Protections}},
  howpublished = {Cryptology ePrint Archive, Paper 2025/1261},
  year         = {2025},
  url          = {https://eprint.iacr.org/2025/1261}
}

@article{deshpande2024sdith,
  author  = {Sanjay Deshpande and James Howe and Jakub Szefer and Dongze Yue},
  title   = {{SDitH in Hardware}},
  journal = {IACR Transactions on Cryptographic Hardware and Embedded Systems},
  volume  = {2024},
  number  = {2},
  pages   = {215--251},
  year    = {2024},
  doi     = {10.46586/tches.v2024.i2.215-251},
  url     = {https://doi.org/10.46586/tches.v2024.i2.215-251}
}

@article{bettaieb2024enabling,
  author  = {Slim Bettaieb and Lo{\"i}c Bidoux and Alessandro Budroni and Marco Palumbi and Lucas Pandolfo Perin},
  title   = {{Enabling PERK and Other MPC-in-the-Head Signatures on Resource-Constrained Devices}},
  journal = {IACR Transactions on Cryptographic Hardware and Embedded Systems},
  volume  = {2024},
  number  = {4},
  pages   = {84--109},
  year    = {2024},
  doi     = {10.46586/tches.v2024.i4.84-109},
  url     = {https://doi.org/10.46586/tches.v2024.i4.84-109}
}

@misc{funk2026hake,
  author       = {Brendan Funk and Tianyou Bao and Lo{\"i}c Bidoux and Jiafeng Xie},
  title        = {{HAKE: Efficient Hardware Accelerator for Key Generation of Post-Quantum Signature Scheme PERK}},
  howpublished = {Cryptology ePrint Archive, Paper 2026/841},
  year         = {2026},
  url          = {https://eprint.iacr.org/2026/841}
}

@misc{RISCVISA2026,
  author       = {{RISC-V International}},
  title        = {{The RISC-V Instruction Set Manual, Volume I: Unprivileged Architecture}},
  year         = {2026},
  note         = {{Version 20260120}},
  howpublished = {\url{https://docs.riscv.org/reference/isa/v20260120/index.html}}
}

@misc{KannwischerEtAlPQM42024,
  author       = {Matthias J. Kannwischer and Markus Krausz and Richard Petri and Shang-Yi Yang},
  title        = {{{pqm4}: Benchmarking NIST Additional Post-Quantum Signature Schemes on Microcontrollers}},
  howpublished = {Cryptology ePrint Archive, Paper 2024/112},
  year         = {2024},
  url          = {https://eprint.iacr.org/2024/112}
}

@techreport{nist2023aes,
  author      = {{National Institute of Standards and Technology}},
  title       = {{Advanced Encryption Standard (AES)}},
  institution = {{National Institute of Standards and Technology}},
  address     = {Gaithersburg, MD},
  type        = {{Federal Information Processing Standards Publication}},
  number      = {197-upd1},
  year        = {2023},
  month       = may,
  doi         = {10.6028/NIST.FIPS.197-upd1},
  url         = {https://doi.org/10.6028/NIST.FIPS.197-upd1}
}

@techreport{nist2015fips202,
  author      = {{National Institute of Standards and Technology}},
  title       = {{SHA-3 Standard: Permutation-Based Hash and Extendable-Output Functions}},
  institution = {{National Institute of Standards and Technology}},
  address     = {Gaithersburg, MD},
  type        = {{Federal Information Processing Standards Publication}},
  number      = {202},
  year        = {2015},
  month       = aug,
  doi         = {10.6028/NIST.FIPS.202},
  url         = {https://doi.org/10.6028/NIST.FIPS.202}
}

@inproceedings{swann2021reconfigurable,
  author    = {Ryan Swann and James E. Stine},
  title     = {{A Reconfigurable Architecture for Improvement and Optimization of Advanced Encryption Standard Hardware}},
  booktitle = {2021 55th Asilomar Conference on Signals, Systems, and Computers},
  pages     = {1181--1185},
  publisher = {IEEE},
  year      = {2021},
  doi       = {10.1109/IEEECONF53345.2021.9723104},
  url       = {https://doi.org/10.1109/IEEECONF53345.2021.9723104}
}

@inproceedings{carril2026pqcuark,
  author    = {Xavier Carril and Alicia Manuel Pasoot and Emanuele Parisi and
               Oriol Farr{\`a}s and Carlos Andr{\'e}s Lara-Ni{\~n}o and Miquel Moret{\'o}},
  title     = {{PQCUARK: A Scalar RISC-V ISA Extension for ML-KEM and ML-DSA}},
  booktitle = {2026 Design, Automation \& Test in Europe Conference (DATE)},
  pages     = {1--7},
  publisher = {IEEE},
  year      = {2026},
  doi       = {10.23919/DATE69613.2026.11539512},
  url       = {https://doi.org/10.23919/DATE69613.2026.11539512}
}

@inproceedings{saarinen2024sloth,
  author    = {Markku-Juhani O. Saarinen},
  title     = {{Accelerating SLH-DSA by Two Orders of Magnitude with a Single Hash Unit}},
  booktitle = {Advances in Cryptology -- CRYPTO 2024},
  series    = {Lecture Notes in Computer Science},
  volume    = {14920},
  pages     = {276--304},
  publisher = {Springer},
  year      = {2024},
  doi       = {10.1007/978-3-031-68376-3_9},
  url       = {https://doi.org/10.1007/978-3-031-68376-3_9}
}

@article{karl2024riscv,
  author  = {Patrick Karl and Jonas Schupp and Tim Fritzmann and Georg Sigl},
  title   = {{Post-Quantum Signatures on RISC-V with Hardware Acceleration}},
  journal = {ACM Transactions on Embedded Computing Systems},
  volume  = {23},
  number  = {2},
  pages   = {30:1--30:23},
  year    = {2024},
  month   = mar,
  doi     = {10.1145/3579092},
  url     = {https://doi.org/10.1145/3579092}
}

@article{karl2025performance,
  author    = {Patrick Karl and Jonas Schupp and Georg Sigl},
  title     = {{Performance and Communication Cost of Hardware Accelerators for Hashing in Post-Quantum Cryptography}},
  journal   = {ACM Transactions on Embedded Computing Systems},
  volume    = {24},
  number    = {5},
  articleno = {66},
  pages     = {66:1--66:31},
  year      = {2025},
  month     = sep,
  doi       = {10.1145/3676965},
  url       = {https://doi.org/10.1145/3676965}
}

@article{dolmeta2026horcrux,
  author        = {Alessandra Dolmeta and Valeria Piscopo and Michael Hutter and Maurizio Martina and Guido Masera},
  title         = {{HORCRUX: A Complete PQC RISC-V eXtension Architecture}},
  journal       = {arXiv preprint arXiv:2607.13939},
  year          = {2026},
  eprint        = {2607.13939},
  archivePrefix = {arXiv},
  url           = {https://arxiv.org/abs/2607.13939}
}

@inproceedings{franklin1992communication,
  author    = {Matthew K. Franklin and Moti Yung},
  title     = {{Communication Complexity of Secure Computation (Extended Abstract)}},
  booktitle = {Proceedings of the Twenty-Fourth Annual ACM Symposium on Theory of Computing},
  pages     = {699--710},
  publisher = {ACM},
  year      = {1992},
  doi       = {10.1145/129712.129780},
  url       = {https://doi.org/10.1145/129712.129780}
}

\end{document}